\documentclass[conference,10pt]{IEEEtran}

\IEEEoverridecommandlockouts

\usepackage[utf8]{inputenc}
\usepackage{cite}
\usepackage{amsmath,amsfonts,amssymb,amsthm}
\usepackage{algorithm}
\usepackage{algorithmic}
\usepackage{graphicx}
\usepackage{textcomp}
\usepackage{xcolor}
\usepackage{microtype}
\usepackage{balance}
\usepackage[caption=false,font=footnotesize]{subfig}
\usepackage{booktabs}
\usepackage{multirow}
\usepackage{bm}
\usepackage{hyperref}

\usepackage{tikz}
\usepackage{times}
\usepackage{pgfplots}

\pgfplotsset{compat=newest}

\usepgfplotslibrary{groupplots}
\usepgfplotslibrary{dateplot}

\tikzset{every picture/.append style={line join=round,line cap=round,}}
\tikzstyle{solid}=[dash pattern=]
\tikzstyle{dotted}=[dash pattern=on 0.0\pgflinewidth off 2.5\pgflinewidth]
\tikzstyle{dashed}=[dash pattern=on 4.0\pgflinewidth off 3.0\pgflinewidth]
\tikzstyle{dashdotted}=[dash pattern=on 0.0\pgflinewidth off 2.0\pgflinewidth on 3.0\pgflinewidth off 2.0\pgflinewidth]
\pgfplotsset{every tick label/.append style={font=\Large}}
\pgfplotsset{every axis legend/.append style={font=\large}}
\pgfplotsset{every axis label/.append style={font=\Large}}
\pgfplotsset{every axis/.append style={width=7cm,height=4cm}}

\allowdisplaybreaks 
\makeatletter
\newcommand\fs@betterruled{%
  \def\@fs@cfont{\bfseries}\let\@fs@capt\floatc@ruled
  \def\@fs@pre{\vspace*{5pt}\hrule height.8pt depth0pt \kern2pt}%
  \def\@fs@post{\kern2pt\hrule\relax}%
  \def\@fs@mid{\kern2pt\hrule\kern2pt}%
  \let\@fs@iftopcapt\iftrue}
\floatstyle{betterruled}
\restylefloat{algorithm}
\makeatother

\usepackage{amssymb}
\usepackage{amsfonts}
\usepackage{mathrsfs}
\usepackage{xspace}
\usepackage{bm}
\usepackage{upgreek}

\newcommand{\safemath}[2]{\newcommand{#1}{\ensuremath{#2}\xspace}}

\safemath{\bma}{\mathbf{a}}
\safemath{\bmb}{\mathbf{b}}
\safemath{\bmc}{\mathbf{c}}
\safemath{\bmd}{\mathbf{d}}
\safemath{\bme}{\mathbf{e}}
\safemath{\bmf}{\mathbf{f}}
\safemath{\bmg}{\mathbf{g}}
\safemath{\bmh}{\mathbf{h}}
\safemath{\bmi}{\mathbf{i}}
\safemath{\bmj}{\mathbf{j}}
\safemath{\bmk}{\mathbf{k}}
\safemath{\bml}{\mathbf{l}}
\safemath{\bmm}{\mathbf{m}}
\safemath{\bmn}{\mathbf{n}}
\safemath{\bmo}{\mathbf{o}}
\safemath{\bmp}{\mathbf{p}}
\safemath{\bmq}{\mathbf{q}}
\safemath{\bmr}{\mathbf{r}}
\safemath{\bms}{\mathbf{s}}
\safemath{\bmt}{\mathbf{t}}
\safemath{\bmu}{\mathbf{u}}
\safemath{\bmv}{\mathbf{v}}
\safemath{\bmw}{\mathbf{w}}
\safemath{\bmx}{\mathbf{x}}
\safemath{\bmy}{\mathbf{y}}
\safemath{\bmz}{\mathbf{z}}
\safemath{\bmzero}{\mathbf{0}}
\safemath{\bmone}{\mathbf{1}}

\bmdefine{\biad}{a}
\bmdefine{\bibd}{b}
\bmdefine{\bicd}{c}
\bmdefine{\bidd}{d}
\bmdefine{\bied}{e}
\bmdefine{\bifd}{f}
\bmdefine{\bigd}{g}
\bmdefine{\bihd}{h}
\bmdefine{\biid}{i}
\bmdefine{\bijd}{j}
\bmdefine{\bikd}{k}
\bmdefine{\bild}{l}
\bmdefine{\bimd}{m}
\bmdefine{\bind}{n}
\bmdefine{\biod}{o}
\bmdefine{\bipd}{p}
\bmdefine{\biqd}{q}
\bmdefine{\bird}{r}
\bmdefine{\bisd}{s}
\bmdefine{\bitd}{t}
\bmdefine{\biud}{u}
\bmdefine{\bivd}{v}
\bmdefine{\biwd}{w}
\bmdefine{\bixd}{x}
\bmdefine{\biyd}{y}
\bmdefine{\bizd}{z}

\bmdefine{\bixid}{\xi}
\bmdefine{\bilambdad}{\lambda}
\bmdefine{\bimud}{\mu}
\bmdefine{\bithetad}{\theta}
\bmdefine{\biphid}{\phi}
\bmdefine{\bideltad}{\delta}

\safemath{\bmia}{\biad}
\safemath{\bmib}{\bibd}
\safemath{\bmic}{\bicd}
\safemath{\bmid}{\bidd}
\safemath{\bmie}{\bied}
\safemath{\bmif}{\bifd}
\safemath{\bmig}{\bigd}
\safemath{\bmih}{\bihd}
\safemath{\bmii}{\biid}
\safemath{\bmij}{\bijd}
\safemath{\bmik}{\bikd}
\safemath{\bmil}{\bild}
\safemath{\bmim}{\bimd}
\safemath{\bmin}{\bind}
\safemath{\bmio}{\biod}
\safemath{\bmip}{\bipd}
\safemath{\bmiq}{\biqd}
\safemath{\bmir}{\bird}
\safemath{\bmis}{\bisd}
\safemath{\bmit}{\bitd}
\safemath{\bmiu}{\biud}
\safemath{\bmiv}{\bivd}
\safemath{\bmiw}{\biwd}
\safemath{\bmix}{\bixd}
\safemath{\bmiy}{\biyd}
\safemath{\bmiz}{\bizd}

\safemath{\bmxi}{\bixid}
\safemath{\bmlambda}{\bilambdad}
\safemath{\bmmu}{\bimud}
\safemath{\bmtheta}{\bithetad}
\safemath{\bmphi}{\biphid}
\safemath{\bmdelta}{\bideltad}

\safemath{\bA}{\mathbf{A}}
\safemath{\bB}{\mathbf{B}}
\safemath{\bC}{\mathbf{C}}
\safemath{\bD}{\mathbf{D}}
\safemath{\bE}{\mathbf{E}}
\safemath{\bF}{\mathbf{F}}
\safemath{\bG}{\mathbf{G}}
\safemath{\bH}{\mathbf{H}}
\safemath{\bI}{\mathbf{I}}
\safemath{\bJ}{\mathbf{J}}
\safemath{\bK}{\mathbf{K}}
\safemath{\bL}{\mathbf{L}}
\safemath{\bM}{\mathbf{M}}
\safemath{\bN}{\mathbf{N}}
\safemath{\bO}{\mathbf{O}}
\safemath{\bP}{\mathbf{P}}
\safemath{\bQ}{\mathbf{Q}}
\safemath{\bR}{\mathbf{R}}
\safemath{\bS}{\mathbf{S}}
\safemath{\bT}{\mathbf{T}}
\safemath{\bU}{\mathbf{U}}
\safemath{\bV}{\mathbf{V}}
\safemath{\bW}{\mathbf{W}}
\safemath{\bX}{\mathbf{X}}
\safemath{\bY}{\mathbf{Y}}
\safemath{\bZ}{\mathbf{Z}}

\safemath{\bZero}{\mathbf{0}}
\safemath{\bOne}{\mathbf{1}}
\safemath{\bDelta}{\mathbf{\Delta}}
\safemath{\bLambda}{\mathbf{\UpLambda}}
\safemath{\bPhi}{\mathbf{\Phi}}
\safemath{\bPsi}{\mathbf{\Psi}}
\safemath{\bSigma}{\mathbf{\Upsigma}}
\safemath{\bOmega}{\mathbf{\Upomega}}
\safemath{\bTheta}{\mathbf{\Uptheta}}

\bmdefine{\biAd}{A}
\bmdefine{\biBd}{B}
\bmdefine{\biCd}{C}
\bmdefine{\biDd}{D}
\bmdefine{\biEd}{E}
\bmdefine{\biFd}{F}
\bmdefine{\biGd}{G}
\bmdefine{\biHd}{H}
\bmdefine{\biId}{I}
\bmdefine{\biJd}{J}
\bmdefine{\biKd}{K}
\bmdefine{\biLd}{L}
\bmdefine{\biMd}{M}
\bmdefine{\biOd}{N}
\bmdefine{\biPd}{O}
\bmdefine{\biQd}{P}
\bmdefine{\biRd}{R}
\bmdefine{\biSd}{S}
\bmdefine{\biTd}{T}
\bmdefine{\biUd}{U}
\bmdefine{\biVd}{V}
\bmdefine{\biWd}{W}
\bmdefine{\biXd}{X}
\bmdefine{\biYd}{Y}
\bmdefine{\biZd}{Z}

\bmdefine{\biDelta}{\Delta}
\bmdefine{\biLambda}{\Lambda}
\bmdefine{\biPhi}{\Phi}
\bmdefine{\biSigma}{\Sigma}
\bmdefine{\biOmega}{\Omega}
\bmdefine{\biTheta}{\Theta}

\safemath{\bimA}{\biAd}
\safemath{\bimB}{\biBd}
\safemath{\bimC}{\biCd}
\safemath{\bimD}{\biDd}
\safemath{\bimE}{\biEd}
\safemath{\bimF}{\biFd}
\safemath{\bimG}{\biGd}
\safemath{\bimH}{\biHd}
\safemath{\bimI}{\biId}
\safemath{\bimJ}{\biJd}
\safemath{\bimK}{\biKd}
\safemath{\bimL}{\biLd}
\safemath{\bimM}{\biMd}
\safemath{\bimN}{\biNd}
\safemath{\bimO}{\biOd}
\safemath{\bimP}{\biPd}
\safemath{\bimQ}{\biQd}
\safemath{\bimR}{\biRd}
\safemath{\bimS}{\biSd}
\safemath{\bimT}{\biTd}
\safemath{\bimU}{\biUd}
\safemath{\bimV}{\biVd}
\safemath{\bimW}{\biWd}
\safemath{\bimX}{\biXd}
\safemath{\bimY}{\biYd}
\safemath{\bimZ}{\biZd}

\safemath{\bimDelta}{\biDelta}
\safemath{\bimLambda}{\biLambda}
\safemath{\bimPhi}{\biPhi}
\safemath{\bimSigma}{\biSigma}
\safemath{\bimOmega}{\biOmega}
\safemath{\bimTheta}{\biTheta}

\safemath{\setA}{\mathcal{A}}
\safemath{\setB}{\mathcal{B}}
\safemath{\setC}{\mathcal{C}}
\safemath{\setD}{\mathcal{D}}
\safemath{\setE}{\mathcal{E}}
\safemath{\setF}{\mathcal{F}}
\safemath{\setG}{\mathcal{G}}
\safemath{\setH}{\mathcal{H}}
\safemath{\setI}{\mathcal{I}}
\safemath{\setJ}{\mathcal{J}}
\safemath{\setK}{\mathcal{K}}
\safemath{\setL}{\mathcal{L}}
\safemath{\setM}{\mathcal{M}}
\safemath{\setN}{\mathcal{N}}
\safemath{\setO}{\mathcal{O}}
\safemath{\setP}{\mathcal{P}}
\safemath{\setQ}{\mathcal{Q}}
\safemath{\setR}{\mathcal{R}}
\safemath{\setS}{\mathcal{S}}
\safemath{\setT}{\mathcal{T}}
\safemath{\setU}{\mathcal{U}}
\safemath{\setV}{\mathcal{V}}
\safemath{\setW}{\mathcal{W}}
\safemath{\setX}{\mathcal{X}}
\safemath{\setY}{\mathcal{Y}}
\safemath{\setZ}{\mathcal{Z}}
\safemath{\emptySet}{\varnothing}

\safemath{\colA}{\mathscr{A}}
\safemath{\colB}{\mathscr{B}}
\safemath{\colC}{\mathscr{C}}
\safemath{\colD}{\mathscr{D}}
\safemath{\colE}{\mathscr{E}}
\safemath{\colF}{\mathscr{F}}
\safemath{\colG}{\mathscr{G}}
\safemath{\colH}{\mathscr{H}}
\safemath{\colI}{\mathscr{I}}
\safemath{\colJ}{\mathscr{J}}
\safemath{\colK}{\mathscr{K}}
\safemath{\colL}{\mathscr{L}}
\safemath{\colM}{\mathscr{M}}
\safemath{\colN}{\mathscr{N}}
\safemath{\colO}{\mathscr{O}}
\safemath{\colP}{\mathscr{P}}
\safemath{\colQ}{\mathscr{Q}}
\safemath{\colR}{\mathscr{R}}
\safemath{\colS}{\mathscr{S}}
\safemath{\colT}{\mathscr{T}}
\safemath{\colU}{\mathscr{U}}
\safemath{\colV}{\mathscr{V}}
\safemath{\colW}{\mathscr{W}}
\safemath{\colX}{\mathscr{X}}
\safemath{\colY}{\mathscr{Y}}
\safemath{\colZ}{\mathscr{Z}}

\safemath{\opA}{\mathbb{A}}
\safemath{\opB}{\mathbb{B}}
\safemath{\opC}{\mathbb{C}}
\safemath{\opD}{\mathbb{D}}
\safemath{\opE}{\mathbb{E}}
\safemath{\opF}{\mathbb{F}}
\safemath{\opG}{\mathbb{G}}
\safemath{\opH}{\mathbb{H}}
\safemath{\opI}{\mathbb{I}}
\safemath{\opJ}{\mathbb{J}}
\safemath{\opK}{\mathbb{K}}
\safemath{\opL}{\mathbb{L}}
\safemath{\opM}{\mathbb{M}}
\safemath{\opN}{\mathbb{N}}
\safemath{\opO}{\mathbb{O}}
\safemath{\opP}{\mathbb{P}}
\safemath{\opQ}{\mathbb{Q}}
\safemath{\opR}{\mathbb{R}}
\safemath{\opS}{\mathbb{S}}
\safemath{\opT}{\mathbb{T}}
\safemath{\opU}{\mathbb{U}}
\safemath{\opV}{\mathbb{V}}
\safemath{\opW}{\mathbb{W}}
\safemath{\opX}{\mathbb{X}}
\safemath{\opY}{\mathbb{Y}}
\safemath{\opZ}{\mathbb{Z}}
\safemath{\opZero}{\mathbb{O}}
\safemath{\identityop}{\opI}

\safemath{\veca}{\bma}
\safemath{\vecb}{\bmb}
\safemath{\vecc}{\bmc}
\safemath{\vecd}{\bmd}
\safemath{\vece}{\bme}
\safemath{\vecf}{\bmf}
\safemath{\vecg}{\bmg}
\safemath{\vech}{\bmh}
\safemath{\veci}{\bmi}
\safemath{\vecj}{\bmj}
\safemath{\veck}{\bmk}
\safemath{\vecl}{\bml}
\safemath{\vecm}{\bmm}
\safemath{\vecn}{\bmn}
\safemath{\veco}{\bmo}
\safemath{\vecp}{\bmp}
\safemath{\vecq}{\bmq}
\safemath{\vecr}{\bmr}
\safemath{\vecs}{\bms}
\safemath{\vect}{\bmt}
\safemath{\vecu}{\bmu}
\safemath{\vecv}{\bmv}
\safemath{\vecw}{\bmw}
\safemath{\vecx}{\bmx}
\safemath{\vecy}{\bmy}
\safemath{\vecz}{\bmz}

\safemath{\veczero}{\bmzero}
\safemath{\vecone}{\bmone}
\safemath{\vecxi}{\bmxi}
\safemath{\veclambda}{\bmlambda}
\safemath{\vecmu}{\bmmu}
\safemath{\vectheta}{\bmtheta}
\safemath{\vecphi}{\bmphi}
\safemath{\vecdelta}{\bmdelta}

\safemath{\matA}{\bA}
\safemath{\matB}{\bB}
\safemath{\matC}{\bC}
\safemath{\matD}{\bD}
\safemath{\matE}{\bE}
\safemath{\matF}{\bF}
\safemath{\matG}{\bG}
\safemath{\matH}{\bH}
\safemath{\matI}{\bI}
\safemath{\matJ}{\bJ}
\safemath{\matK}{\bK}
\safemath{\matL}{\bL}
\safemath{\matM}{\bM}
\safemath{\matN}{\bN}
\safemath{\matO}{\bO}
\safemath{\matP}{\bP}
\safemath{\matQ}{\bQ}
\safemath{\matR}{\bR}
\safemath{\matS}{\bS}
\safemath{\matT}{\bT}
\safemath{\matU}{\bU}
\safemath{\matV}{\bV}
\safemath{\matW}{\bW}
\safemath{\matX}{\bX}
\safemath{\matY}{\bY}
\safemath{\matZ}{\bZ}
\safemath{\matzero}{\bmzero}

\safemath{\matDelta}{\bDelta}
\safemath{\matLambda}{\bLambda}
\safemath{\matPhi}{\bPhi}
\safemath{\matSigma}{\bSigma}
\safemath{\matOmega}{\bOmega}
\safemath{\matTheta}{\bTheta}

\safemath{\matidentity}{\matI}
\safemath{\matone}{\matO}

\safemath{\rnda}{A}
\safemath{\rndb}{B}
\safemath{\rndc}{C}
\safemath{\rndd}{D}
\safemath{\rnde}{E}
\safemath{\rndf}{F}
\safemath{\rndg}{G}
\safemath{\rndh}{H}
\safemath{\rndi}{I}
\safemath{\rndj}{J}
\safemath{\rndk}{K}
\safemath{\rndl}{L}
\safemath{\rndm}{M}
\safemath{\rndn}{N}
\safemath{\rndo}{O}
\safemath{\rndp}{P}
\safemath{\rndq}{Q}
\safemath{\rndr}{R}
\safemath{\rnds}{S}
\safemath{\rndt}{T}
\safemath{\rndu}{U}
\safemath{\rndv}{V}
\safemath{\rndw}{W}
\safemath{\rndx}{X}
\safemath{\rndy}{Y}
\safemath{\rndz}{Z}

\safemath{\rveca}{\bimA}
\safemath{\rvecb}{\bimB}
\safemath{\rvecc}{\bimC}
\safemath{\rvecd}{\bimD}
\safemath{\rvece}{\bimE}
\safemath{\rvecf}{\bimF}
\safemath{\rvecg}{\bimG}
\safemath{\rvech}{\bimH}
\safemath{\rveci}{\bimI}
\safemath{\rvecj}{\bimJ}
\safemath{\rveck}{\bimK}
\safemath{\rvecl}{\bimL}
\safemath{\rvecm}{\bimM}
\safemath{\rvecn}{\bimN}
\safemath{\rveco}{\bomO}
\safemath{\rvecp}{\bimP}
\safemath{\rvecq}{\bimQ}
\safemath{\rvecr}{\bimR}
\safemath{\rvecs}{\bimS}
\safemath{\rvect}{\bimT}
\safemath{\rvecu}{\bimU}
\safemath{\rvecv}{\bimV}
\safemath{\rvecw}{\bimW}
\safemath{\rvecx}{\bimX}
\safemath{\rvecy}{\bimY}
\safemath{\rvecz}{\bimZ}

\safemath{\rvecxi}{\bmxi}
\safemath{\rveclambda}{\bmlambda}
\safemath{\rvecmu}{\bmmu}
\safemath{\rvectheta}{\bmtheta}
\safemath{\rvecphi}{\bmphi}

\safemath{\rmatA}{\bimA}
\safemath{\rmatB}{\bimB}
\safemath{\rmatC}{\bimC}
\safemath{\rmatD}{\bimD}
\safemath{\rmatE}{\bimE}
\safemath{\rmatF}{\bimF}
\safemath{\rmatG}{\bimG}
\safemath{\rmatH}{\bimH}
\safemath{\rmatI}{\bimI}
\safemath{\rmatJ}{\bimJ}
\safemath{\rmatK}{\bimK}
\safemath{\rmatL}{\bimL}
\safemath{\rmatM}{\bimM}
\safemath{\rmatN}{\bimN}
\safemath{\rmatO}{\bimO}
\safemath{\rmatP}{\bimP}
\safemath{\rmatQ}{\bimQ}
\safemath{\rmatR}{\bimR}
\safemath{\rmatS}{\bimS}
\safemath{\rmatT}{\bimT}
\safemath{\rmatU}{\bimU}
\safemath{\rmatV}{\bimV}
\safemath{\rmatW}{\bimW}
\safemath{\rmatX}{\bimX}
\safemath{\rmatY}{\bimY}
\safemath{\rmatZ}{\bimZ}

\safemath{\rmatDelta}{\bimDelta}
\safemath{\rmatLambda}{\bimLambda}
\safemath{\rmatPhi}{\bimPhi}
\safemath{\rmatSigma}{\bimSigma}
\safemath{\rmatOmega}{\bimOmega}
\safemath{\rmatTheta}{\bimTheta}

\usepackage{amssymb}
\usepackage{amsfonts}
\usepackage{mathrsfs}
\usepackage{xspace}
\usepackage{bm}
\usepackage{fancyref}
\usepackage{textcomp}

\usepackage{multirow}
\usepackage{stmaryrd}

\newenvironment{textbmatrix}{	\setlength{\arraycolsep}{2.5pt}%
								\big[\begin{matrix}}{\end{matrix}\big]%
								\raisebox{0.08ex}{\vphantom{M}}}

\def\be{\begin{equation}}
\def\ee{\end{equation}}
\def\een{\nonumber \end{equation}}
\def\mat{\begin{bmatrix}}
\def\emat{\end{bmatrix}}
\def\btm{\begin{textbmatrix}}
\def\etm{\end{textbmatrix}}

\def\ba#1\ea{\begin{align}#1\end{align}}
\def\bas#1\eas{\begin{align*}#1\end{align*}}
\def\bs#1\es{\begin{split}#1\end{split}} 
\def\bg#1\eg{\begin{gather}#1\end{gather}}
\def\bml#1\eml{\begin{multline}#1\end{multline}}
\def\bi#1\ei{\begin{itemize}#1\end{itemize}}

\safemath{\dirac}{\delta}					
\safemath{\krond}{\dirac}					

\safemath{\upto}{\uparrow}
\safemath{\downto}{\downarrow}
\safemath{\iu}{j}							
\safemath{\ev}{\lambda}						
\safemath{\hilseqspace}{l^{2}}				
\newcommand{\banachfunspace}[1]{\setL^{#1}}	
\safemath{\hilfunspace}{\banachfunspace{2}}	

\safemath{\SNR}{\textsf{SNR}} 				
\safemath{\PAR}{\textsf{PAR}} 				
\safemath{\No}{N_0}							
\safemath{\Es}{E_s}							
\safemath{\Eb}{E_b}							
\safemath{\EbNo}{\frac{\Eb}{\No}}
\safemath{\EsNo}{\frac{\Es}{\No}}

\DeclareMathOperator{\CHop}{\ensuremath{\opH}} 
\safemath{\tvir}{\rndh_{\CHop}}				
\safemath{\tvtf}{\rndl_{\CHop}}				
\safemath{\spf}{\rnds_{\CHop}}				
\safemath{\bff}{H_{\CHop}}					

\safemath{\ircf}{r_{h}}						
\safemath{\tftvcf}{r_{s}}					
\safemath{\tfcf}{r_{l}}						
\safemath{\bfcf}{r_{H}}						

\safemath{\tcorr}{c_h}						
\safemath{\scf}{c_{s}}						
\safemath{\tfcorr}{c_{l}}					
\safemath{\fcorr}{c_{H}}						

\safemath{\mi}{I}							
\safemath{\capacity}{C}						

\safemath{\normal}{\mathcal{N}}			
\safemath{\jpg}{\mathcal{CN}}			
\safemath{\mchain}{\leftrightarrow}		

\safemath{\dB}{\,\mathrm{dB}}
\safemath{\dBm}{\,\mathrm{dBm}}
\safemath{\Hz}{\,\mathrm{Hz}}
\safemath{\kHz}{\,\mathrm{kHz}}
\safemath{\MHz}{\,\mathrm{MHz}}
\safemath{\GHz}{\,\mathrm{GHz}}
\safemath{\s}{\,\mathrm{s}}
\safemath{\ms}{\,\mathrm{ms}}
\safemath{\mus}{\,\mathrm{\text{\textmu}s}}
\safemath{\ns}{\,\mathrm{ns}}
\safemath{\ps}{\,\mathrm{ps}}
\safemath{\meter}{\,\mathrm{m}}
\safemath{\mm}{\,\mathrm{mm}}
\safemath{\cm}{\,\mathrm{cm}}
\safemath{\m}{\,\mathrm{m}}
\safemath{\W}{\,\mathrm{W}}
\safemath{\mW}{\, \mathrm{mW}}
\safemath{\J}{\,\mathrm{J}}
\safemath{\K}{\,\mathrm{K}}
\safemath{\bit}{\,\mathrm{bit}}
\safemath{\nat}{\,\mathrm{nat}}

\safemath{\define}{\triangleq}			

\safemath{\equivalent}{\sim}
\safemath{\distas}{\sim}					
\safemath{\sdiff}{\Delta}				

\safemath{\reals}{\mathbb{R}}
\safemath{\positivereals}{\reals_{+}}
\safemath{\integers}{\mathbb{Z}}
\safemath{\posint}{\integers_{+}}
\safemath{\naturals}{\mathbb{N}}
\safemath{\posnaturals}{\naturals_{+}}
\safemath{\complexset}{\mathbb{C}}
\safemath{\rationals}{\mathbb{Q}}

\newcommand*{\fancyrefapplabelprefix}{app}		
\newcommand*{\fancyrefthmlabelprefix}{thm}		
\newcommand*{\fancyreflemlabelprefix}{lem}		
\newcommand*{\fancyrefcorlabelprefix}{cor}		
\newcommand*{\fancyrefdeflabelprefix}{def}		
\newcommand*{\fancyrefproplabelprefix}{prop}	
\newcommand*{\fancyrefobslabelprefix}{obs}		
\newcommand*{\fancyrefalglabelprefix}{alg}		
\newcommand*{\fancyrefasmlabelprefix}{asm}	    
\newcommand*{\fancyreftbllabelprefix}{tbl}	    

\frefformat{vario}{\fancyrefseclabelprefix}{Section~#1}
\frefformat{vario}{\fancyrefthmlabelprefix}{Theorem~#1}
\frefformat{vario}{\fancyreflemlabelprefix}{Lemma~#1}
\frefformat{vario}{\fancyrefcorlabelprefix}{Corollary~#1}
\frefformat{vario}{\fancyrefdeflabelprefix}{Definition~#1}
\frefformat{vario}{\fancyrefobslabelprefix}{Observation~#1}
\frefformat{vario}{\fancyrefasmlabelprefix}{Assumption~#1}
\frefformat{vario}{\fancyreffiglabelprefix}{Figure~#1}
\frefformat{vario}{\fancyrefapplabelprefix}{Appendix~#1} 
\frefformat{vario}{\fancyrefproplabelprefix}{Proposition~#1}
\frefformat{vario}{\fancyrefalglabelprefix}{Algorithm~#1}
\frefformat{vario}{\fancyrefeqlabelprefix}{(#1)}
\frefformat{vario}{\fancyreftbllabelprefix}{Table~#1}

\safemath{\dictab}{[\,\dicta\,\,\dictb\,]}

\safemath{\ysig}{\bmy}
\safemath{\ysighat}{\hat{\ysig}}
\safemath{\ysigdim}{M}
\safemath{\xsig}{\bmx}
\safemath{\xsigdim}{N}
\safemath{\nx}{n_x}
\safemath{\zsig}{\bmz}
\safemath{\zsigdim}{\ysigdim}
\safemath{\rsig}{\bmr}
\safemath{\Adict}{\bA}
\safemath{\Adicttilde}{\widetilde{\Adict}}
\safemath{\Adictdim}{\outputdim\times\xsigdim}
\safemath{\avec}{\bma}
\safemath{\avectilde}{\tilde{\avec}}
\safemath{\Bdict}{\bB}
\safemath{\Bdicttilde}{\widetilde{\Bdict}}
\safemath{\Cdict}{\bC}
\safemath{\cvec}{\bmc}
\safemath{\Ddict}{\bD}
\safemath{\Ddictdim}{\ysigdim\times\xsigdim}
\safemath{\dvec}{\bmd}
\safemath{\Ddicttilde}{\widetilde{\bD}}
\safemath{\Bonb}{\bB}
\safemath{\bvec}{\bmb}
\safemath{\Bonbdim}{\ysigdim\times\ysigdim}
\safemath{\noise}{\bmn}
\safemath{\noisedim}{\ysigim}
\safemath{\err}{\bme}
\safemath{\errdim}{\ysigdim}
\safemath{\errset}{\setE}
\safemath{\nerr}{n_e}
\safemath{\delop}{\bP_\errset}
\safemath{\delopc}{\bP_{{\errset}^c}}

\safemath{\cplxi}{\imath}
\safemath{\cplxj}{\jmath}

\safemath{\dict}{\matD}
\safemath{\inputdim}{N}		
\safemath{\outputdim}{M}		
\safemath{\sparsity}{S}	
\safemath{\inputdimA}{{N_a}}	
\safemath{\inputdimB}{{N_b}}	
\safemath{\elemA}{{n_a}}	
\safemath{\elemB}{{n_b}}	
\safemath{\resA}{\matR_a}	
\safemath{\resB}{\matR_b}	
\safemath{\subD}{\matS} 
\safemath{\subA}{\matS_a} 
\safemath{\subB}{\matS_b} 
\safemath{\dicta}{\matA} 	
\safemath{\dictb}{\matB} 	
\safemath{\hollowS}{H}
\safemath{\hollowA}{H_a}
\safemath{\hollowB}{H_b}
\safemath{\cross}{Z}
\safemath{\coh}{\mu_d}			
\safemath{\coha}{\mu_a}			
\safemath{\cohb}{\mu_b}			
\safemath{\mubs}{\nu}	
\safemath{\cohm}{\mu_m} 
\safemath{\dictset}{\setD}	
\safemath{\dictsetp}{\dictset(\coh,\coha,\cohb)}	
\safemath{\dictsetgen}{\dictset_\text{gen}}
\safemath{\dictsetgenp}{\dictsetgen(\coh)}
\safemath{\dictsetonb}{\dictset_\text{onb}}
\safemath{\dictsetonbp}{\dictsetonb(\coh)}

\safemath{\leftside}{U}
\safemath{\rightsideA}{R_a}
\safemath{\rightsideB}{R_b}

\safemath{\indexS}{\setI_S} 

\safemath{\na}{n_a}			
\safemath{\nb}{n_b}			
\safemath{\coeffa}{p_i}	
\safemath{\coeffb}{q_j}	
\safemath{\seta}{\setP}		
\safemath{\setb}{\setQ}     
\safemath{\setw}{\setW}	
\safemath{\setz}{\setZ}	
\safemath{\cola}{\veca}		
\safemath{\colb}{\vecb}		
\safemath{\cold}{\vecd}		
\safemath{\inputvec}{\vecx} 	
\safemath{\error}{\vece}	
\safemath{\noiseout}{\vecz} 	
\safemath{\inputvecel}{x}
\safemath{\inputveca}{\vecx_a}
\safemath{\inputvecb}{\vecx_b}
\safemath{\outputvec}{\vecy}	
\safemath{\lambdamin}{\lambda_{\mathrm{min}}}

\safemath{\elltwo}{\ell_2}
\safemath{\ellone}{\ell_1}
\safemath{\ellzero}{\ell_0}
\safemath{\ellinf}{\ell_\infty}
\safemath{\ellinftilde}{\ell_{\widetilde\infty}}
\safemath{\licard}{Z(\coh,\coha,\cohb)}
\safemath{\xsol}{\hat{x}}
\safemath{\xbord}{x_b}		
\safemath{\xstat}{x_s}		
\safemath{\xstatLone}{\tilde{x}_s}
\safemath{\order}{\mathcal{O}} 
\safemath{\scales}{\Theta} 
\safemath{\ones}{\mathbf{1}} 
\safemath{\zeroes}{\mathbf{0}} 
\safemath{\thlone}{\kappa(\coh,\cohb)} 
\safemath{\constoneA}{\delta} 
\safemath{\constoneB}{\epsilon} 
\safemath{\nlarge}{L}				   
\safemath{\sumlarge}{S_\nlarge}
\safemath{\maxlarger}{P_\nlarge}	   
\safemath{\Pzero}{\textrm{P0}}	
\safemath{\Pone}{\textrm{P1}}
\safemath{\vecfir}{\vecw}			 
\safemath{\vecsec}{\vecz}
\safemath{\elvecfir}{w}              
\safemath{\elvecsec}{z}				 
\safemath{\nlargefir}{n}
\safemath{\normout}{\gamma}
\safemath{\auxfun}{h}
\safemath{\supp}{\textrm{supp}}

\safemath{\indexa}{\ell}
\safemath{\indexb}{r}
\safemath{\indexc}{i}
\safemath{\indexd}{j}

\safemath{\project}{P}

\begin{document}
\title{Improving Channel Charting using a \\ Split Triplet Loss and an Inertial Regularizer}
\author{\IEEEauthorblockN{Brian Rappaport$^\text{1}$, Emre Gönültaş$^\text{1}$, Jakob Hoydis$^\text{2}$, Maximilian Arnold$^\text{3}$, \\
Pavan Koteshwar Srinath$^\text{4}$, and Christoph Studer$^\text{5}$\\[0.3cm]}
\thanks{The work of CS was supported in part by ComSenTer, one of six centers in JUMP, a SRC program sponsored by DARPA, and by an ETH Research Grant.
The work of BR, EG, and CS was also supported by the US National Science Foundation (NSF) under grants CNS-1717559 and ECCS-1824379.}
\IEEEauthorblockA{\em $^\text{1}$School of Electrical and Computer Engineering, Cornell University; e-mail: br434@cornell.edu, eg566@cornell.edu\\
$^\text{2}$NVIDIA, France; e-mail: jhoydis@nvidia.com \\
$^\text{3}$Nokia Bell Labs Stuttgart; e-mail: maximilian.arnold@nokia-bell-labs.com \\
$^\text{4}$Nokia Bell Labs France; e-mail: pavan.koteshwar\_srinath@nokia-bell-labs.com \\
$^\text{5}$Department of Information Technology and Electrical Engineering, ETH Z\"urich; e-mail: studer@ethz.ch}
}

\maketitle

\begin{abstract}
Channel charting is an emerging technology that enables self-supervised pseudo-localization of user equipments by performing dimensionality reduction on large channel-state information (CSI) databases that are passively collected at infrastructure base stations or access points. In this paper, we introduce a new dimensionality reduction method specifically designed for channel charting using a novel split triplet loss, which utilizes physical information available during the CSI acquisition process. In addition, we propose a novel regularizer that exploits the physical concept of inertia, which significantly improves the quality of the learned channel charts. We provide an experimental verification of our methods using synthetic and real-world measured CSI datasets, and we demonstrate that our methods are able to outperform the state-of-the-art in channel charting based on the triplet loss. 
\end{abstract}

\IEEEpeerreviewmaketitle


\section{Introduction}

Hardly any area of modern research remains untouched by deep-learning-based machine learning. Wireless communications is no exception; in recent years, machine learning has been considered for many applications, including network routing, data detection, and others \cite{forster2007,sun2019application,jiang2016machine}.
The rapid pace of data collection in modern wireless systems allows for data-driven approaches to address many pressing problems, including localization, the problem of estimating the position of a wireless transmitter given its channel state information (CSI). Wireless localization is a well-studied problem; CSI fingerprinting \cite{zhang2019stack,larsson2015fingerprinting} is one approach, but fingerprinting requires large CSI databases and access to ground-truth location, which is challenging to acquire as the channels change over time. To address such issues, we turn to self-supervised methods, which do not require ground truth location information.

\subsection{Channel Charting: Main Idea}
Of particular interest here is \emph{channel charting} \cite{channelcharting}, a self-supervised framework for pseudo-localization and tracking of user equipments (UEs) using dimensionality reduction in the CSI domain.
This technique learns a charting function over a specific channel, which maps high-dimensional CSI to low-dimensional pseudo-locations. Moreover, the function is learned in an self-supervised manner, without any need for ground truth location information.
Successful channel charting relies on two aspects: (i) carefully-designed CSI features, as CSI measurements are prone to a variety of system and hardware impairments, such as noise, small-scale fading, time synchronization errors, and residual sampling rate and frequency offsets; and (ii) the dimensionality reduction method~\cite{channelcharting, pengzhipaper,ericpaper,huawei2020channelcharting,huawei2020paper}. This paper focuses on the second aspect.

We consider one or more mobile UEs moving in the vicinity of an infrastructure base station (BS) or access point (AP), which regularly transmit pilots that are used to acquire large CSI datasets.
The time-varying information can be used to build a channel chart with data from the same UE but taken at different time instants, rather than CSI measurements at the same time from different transmitters. The additional information that two consecutive samples are temporally adjacent also improves the quality of dimensionality reduction. Similar ideas have been considered in the past, e.g., the SM+ algorithm proposed in~\cite{channelcharting}, but have been limited in scope.

\subsection{Contributions}  

In this paper, we strive to take full advantage of all information available at the BS or AP when acquiring CSI measurements and learning the channel chart, both explicit and implicit. Concretely, we propose a new dimensionality reduction method we call the \emph{split triplet loss}, which takes into account physical properties of the transmitting UEs during CSI acquisition. 
In addition, we augment our loss function using a novel \emph{inertial regularizer}, which further improves the quality of the learned channel charting function.
We demonstrate the efficacy of our methods using a synthetic and a real-world dataset, and compare them to Sammon's mapping from \cite{ericpaper} and the recently-proposed triplet loss approach from~\cite{huawei2020channelcharting}.

\subsection{Notation}
We denote column vectors and matrices by boldface lowercase and capital letters, respectively.  
The $i$th element of a vector $\bma$ is denoted by~$a_i$, the $i$th column of a matrix $\bA$ by~$\bma_i$, and the $(i,j)$th entry of $\bA$ by $A_{ij}$. 
The transpose of $\bA$ is $\bA^T$, the conjugate (Hermitian) transpose is $\bA^H$, and the (element-wise) conjugate is $\bA^*$. We use the Euclidean norm for vectors and Frobenius norm for matrices. 
Element-wise powers on vectors and matrices are denoted using $\circ$, e.g., $(\bA^{\circ 2})_{ij} = A_{ij}^2$.


\section{Channel Charting Basics}
\label{sec:channel_charting}

Channel charting is a self-supervised pseudo-localization framework that applies dimensionality reduction to measured CSI~\cite{channelcharting}.
In brief, channel charting takes a dataset of measured CSI, converts the CSI measurements into CSI features that are resilient to small-scale fading and hardware impairments \cite{huawei2020channelcharting,emrepaper}, and applies dimensionality reduction in order to learn a low-dimensional embedding: the \emph{channel chart}. 
Our goal is to learn a channel chart that preserves local geometry: two nearby charted points should also be close in real space.

\subsection{System Model}
We consider a single-input multiple-output (SIMO) orthogonal frequency-division multiplexing system (OFDM) system in which one or multiple single-antenna UEs transmit pilots to a $B$-antenna BS or AP in time-division duplexing (TDD) fashion. 
While the UEs continuously broadcast pilot sequences for all $W$ OFDM subcarriers, the BS or AP collects the SIMO-OFDM CSI of an UE in the form of a CSI matrix $\bH_i\in\complexset^{B\times W}$, where $i$ refers to the index of the measured CSI. 
In what follows, we assume that we are able to track the CSI associated with a specific UE over time, i.e., we can associate each CSI measurement index $i$ with a particular UE $u$ and corresponding time stamp $t_i$.
The collection of CSI measurements $\{\bH_i\}$ forms the dataset that is used to learn the channel chart. 
In what follows, we furthermore assume that there is an \emph{unknown} ground-truth UE location $\bmx_i\in\reals^{D}$ (with $D$ being either two or three) associated with each measured CSI matrix~$\bH_i$.
\subsection{CSI Feature Generation}
Learning a channel chart directly from the CSI measurement dataset $\{\bH_i\}$ is possible, but has several drawbacks that can be mitigated by carefully-crafted CSI features \cite{channelcharting}. In particular, wireless channels are subject to (i) system impairments, mostly small-scale fading caused by moving objects between the transmitter and receiver, and (ii) hardware impairments, caused by varying transmit power, timing synchronization errors, as well as residual carrier frequency and sampling rate offsets.
Following the work of \cite{huawei2020channelcharting,emrepaper}, we use an autocorrelation-based strategy to extract a feature from a raw CSI matrix~$\bH_i$. 

\subsubsection{Normalization}
We normalize the CSI matrix $\bH_i$ via $\bH_i^{\text{norm}} = {\bH_i}/{\|\bH_i\|}$, which makes the CSI features independent of transmit-side power variations and path loss. 

\subsubsection{Delay-Domain Transform}
We convert the normalized CSI matrix into the delay domain by applying an inverse discrete Fourier transform (DFT) to each row of $\bH_i^{\text{norm}}$. 
Concretely, $\bH_i^{\text{delay}} = \bH_i^{\text{norm}}\bF^H$, where $\bF$ is the $W\times W$ unitary DFT matrix.
In an OFDM system with many subcarriers and a short cyclic prefix length $C$, we can truncate the delay-domain CSI matrix by keeping only the $C$ columns of $\bH_i^{\text{delay}}$ corresponding to the impulse response's nonzero coefficients.

\subsubsection{Beamspace Transform}
We convert the delay-domain CSI matrix into the beamspace domain \cite{seyeed2013beamspace}. With a BS/AP utilizing a 1D uniform linear array (ULA) or a 2D uniform rectangular array (URA), we can apply a 1D or 2D DFT for each delay index. For the 1D case, we compute $\bH_i^{\text{beamspace}} = \bF\bH_i^{\text{delay}}$; for the 2D case, we apply the 2D DFT to each column of $\bH_i^{\text{delay}}$.

\subsubsection{Autocorrelation}
We compute an ``instantaneous'' 2D autocorrelation of the beamspace CSI matrix according to $\bH_i^{\text{acorr}} = \bH_i^{\text{beamspace}}*\bH_i^{\text{beamspace}}$, where $*$ is the discrete-time convolution operator. This step renders our CSI features independent of time synchronization offset and global phase changes \cite{emrepaper}. Note that the 2D autocorrelation of a matrix is rotationally symmetric, so we can truncate the redundant second half of the rows of $\bH^{\text{acorr}}$. 

\subsubsection{Vectorization}
We finish by vectorizing the entry-wise absolute values of the truncated autocorrelation matrix in order to extract our final CSI features: $\bmf_i = \text{vec}(|\bH_i^{\text{acorr}}|)$. 

\subsection{Dimensionality Reduction: Sammon's Mapping}
\label{sec:sammon}

After generating CSI features, the next step is to learn a low-dimensional embedding. Dimensionality reduction is a well-studied problem \cite{maaten_dr} and linear dimensionality reduction, e.g., using principal component analysis (PCA), is widely used for many problems.
Unfortunately, such methods do not work well for channel charting and nonlinear methods have been proposed instead~\cite{channelcharting}. 
Sammon's mapping \cite{sammon} learns an embedding vector~$\hat{\bmx}_i$ for each CSI feature $\bmf_i$ in a low ($D'$) dimensional space.
The embedding $\{\hat{\bmx}_i\}$, the set of points in the channel chart, can be calculated by minimizing the non-convex loss function
\begin{align}
\label{eq:Sammon}
\mathcal{L}_{\text{Sammon}}(\{\hat{\bmx}_i\}) \define \sum_{i,j}\frac{(\|\hat{\bmx}_i - \hat{\bmx}_j\| - \|\bmf_i - \bmf_j\|)^2}{\|\bmf_i - \bmf_j\|},
\end{align}
summed over all pairs of CSI features.
Sammon's mapping attempts to learn an embedding $\{\hat{\bmx}_n\}$ in which pairs of CSI features with small Euclidean distance $\|\bmf_i - \bmf_j\|$ match the embedding distance $\|\hat{\bmx}_i - \hat{\bmx}_j\|$; pairs of points with large feature distances are deweighted.
The loss can be approximately minimized efficiently using gradient-descent methods~\cite{channelcharting}.

Although Sammon's mapping is appealingly simple, it has several drawbacks. It is non-parametric and suffers from the out-of-sample problem: given a new CSI feature $\bmf'$, finding the corresponding point $\bmx'$ in the channel chart requires either re-solving the entire optimization problem or using approximate out-of-sample extensions \cite{out_of_sample}.
In \cite{ericpaper}, the authors address this issue using Siamese neural networks \cite{siamesenets}, two parallel and identical neural networks with shared parameter vector $\bithetad$. They replace the embedded points $\hat{\bmx}_i$ and $\hat{\bmx}_j$ in \fref{eq:Sammon} with the outputs of a neural network $g_\bithetad$ that takes in CSI features and produces points in the channel chart by setting $\hat{\bmx}_i = g_\bithetad(\bmf_i)$ and $\hat{\bmx}_j = g_\bithetad(\bmf_j)$.
It can be formulated using the following loss function:
\begin{align} 
\label{eq:Siamese}
\mathcal{L}_{\text{Siamese}}(\bithetad) \define \sum_{i,j}\frac{(\hat{d}_{ij}^\bithetad - \|\bmf_i - \bmf_j\|)^2}{\|\bmf_i - \bmf_j\|},
\end{align}
where $\hat{d}_{ij}^\bithetad \define \|\hat{x}_i^\bithetad - \hat{x}_j^\bithetad\|$, and the network is optimized over the weights and biases subsumed in the vector $\bithetad$.
We note that the embedding points $\hat{\bmx}_i$ and distances $\hat{d}_{ij}$ are functions of $\bithetad$, but we omit these dependencies for ease of notation.
Unfortunately, solving \fref{eq:Siamese} relies heavily on the precept that the Euclidean distance between two CSI features is an accurate proxy for real pairwise distances. Although distance information is contained within the CSI features, the Euclidean distance is a poor choice to uncover it---a more appropriate CSI feature dissimilarity would allow us to learn more faithful channel charts. 

\subsection{Dimensionality Reduction: Triplet Loss}
\label{sec:triplet}

The triplet loss, originally from \cite{facenet} and applied to channel charting in \cite{huawei2020channelcharting}, avoids Euclidean pairwise distances altogether by learning a dissimilarity directly from the CSI features.
This method is based on the concept of triplets: suppose we have access to a set $\mathcal{T} = \{(i,j,k) : d_{ij} \leq d_{ik}\}$, where $d_{ij} \define \|\bmx_i - \bmx_j\|$ is the true pairwise distance. For each triplet $(i,j,k)$, we call the vector $\bmx_i$ the anchor point, $\bmx_j$ the positive (or close) point, and $\bmx_k$ the negative (or far) point. By comparing the distances from positive and negative points, we can learn an embedding which ensures that the ordering of real pairwise distances is maintained in the channel chart.
As before, this method avoids the out-of-sample problem by learning a function~$g_\bithetad$ that maps CSI features to points in the channel chart. 
Reference \cite{huawei2020channelcharting} proposed the following triplet-based loss function for channel charting:
\begin{align} \label{eq:triplet}
\mathcal{L}_{\text{Triplet}}(\bithetad) \define \sum_{(i,j,k)\in\mathcal{T}}[\hat{d}_{ij}^\bithetad - \hat{d}_{ik}^\bithetad + \lambda]^+.
\end{align}
The function $[x]^+ = \max\{x,0\}$ for $x\in\reals$ is the rectified linear unit (ReLU) and $\lambda\in\reals$ is a user-defined margin parameter that determines the global scale of the learned embedding. 
Minimizing this loss function pulls negative points away from and positive points towards the anchor point, which means the ordering of pairwise distances in the embedding will match the true ordering.
In some ways, this approach replaces one problem with another: feature distances are no longer required but one has to determine which points should be positive or negative with respect to an anchor point. 
In \cite{huawei2020channelcharting}, the authors provide a solution based on sampling instants. Using the sampling rate of CSI measurements, they set a temporal bound on triplets: for a given anchor point, points close in time to it are labelled positive and those far in time to it are labelled negative.

The triplet loss has the following shortcomings: it does not make full use of the information available during CSI acquisition. Its generality makes it hard to attach physical significance to results; in particular, the scale of the embedding is entirely dependent on the parameter $\lambda$ in \fref{eq:triplet}. Additionally, it is difficult to tune this loss to make physical sense; the margin parameter $\lambda$ has no physical significance, and it is difficult to select an appropriate margin value in a principled manner.


\section{Split Triplet Loss with Inertial Regularizer}
\label{sec:new_stuff}
To address these concerns, we propose a novel loss function for channel charting, which we call the \emph{split triplet loss}. We also propose an refined method of selecting triplets and a novel regularizer which encapsulates physical constraints on consecutive positions of a moving UE.

\subsection{Split Triplet Loss}

The triplet loss is most effective when pairwise distances are totally ordered: for any triple $(i,j,k)$, it is possible to determine if $d_{ij} \leq d_{ik}$. For channel charting, pairwise distances are only partially ordered, especially when using the temporal triplet selection method described in \fref{sec:triplet}.
It is generally difficult to determine whether $d_{ij} \leq d_{ik}$ for an arbitrary triple $(i,j,k)$. However, for every anchor index $i$, it is possible to categorize all other points as either positive, negative, or neither. 
Consider what information we have about the triplets, given the temporal selection method of section \fref{sec:triplet}. Let $(i,j,k)\in\mathcal{T}$ be a triplet. If we know the maximum and minimum speeds the UEs are moving in space, $v_{\text{max}}$ and $v_{\text{min}}$, then in $T_c$ seconds, we know that the UE can move no further than $b_{\text{pos}} \define v_{\text{max}}\cdot T_c$ and at least $b_{\text{neg}} \define v_{\text{min}}\cdot T_c$.
If the UE is moving in a straight line, then for the triplet loss, $d_{ij} \leq b_{\text{pos}}$ and $d_{ik} \geq b_{\text{neg}}$; if the maximum and minimum speeds are similar, it is quite likely that $d_{ik} \geq d_{ij}$, as required by the triplet loss. However, this information can be encoded more efficiently using two loss terms. The first term determines whether the positive point is close enough to the anchor; the second determines whether the negative point is far enough away. Instead of using constraints, we formulate these properties as regularizers: the first term is $[\hat{d}_{ij}^\bithetad - b_{\text{pos}}]^+$, which is zero if the distance is less than the bound $b_{\text{pos}}$ and increases linearly if not; the second term is $[b_{\text{neg}} - \hat{d}_{ik}^\bithetad]^+$, which is zero if the distance is greater than $b_{\text{neg}}$ and increases linearly as distance decreases.
The overall loss function, the \emph{split triplet loss}, is the sum of these two regularizers:
\begin{align} \label{eq:splittriplet}
\mathcal{L}_{\text{Split\,Triplet}}(\bithetad) \define \sum_{(i,j,k)\in\mathcal{T}}[\hat{d}_{ij}^\bithetad - b_{\text{pos}}]^+ + [b_{\text{neg}} - \hat{d}_{ik}^\bithetad]^+.
\end{align}
This loss function considers physical constraints from the UE movement and measurement process. In contrast, the basic triplet loss does not reflect concepts like speed and does not use available time-stamp information during training.

\pgfplotsset{grid style=dotted}

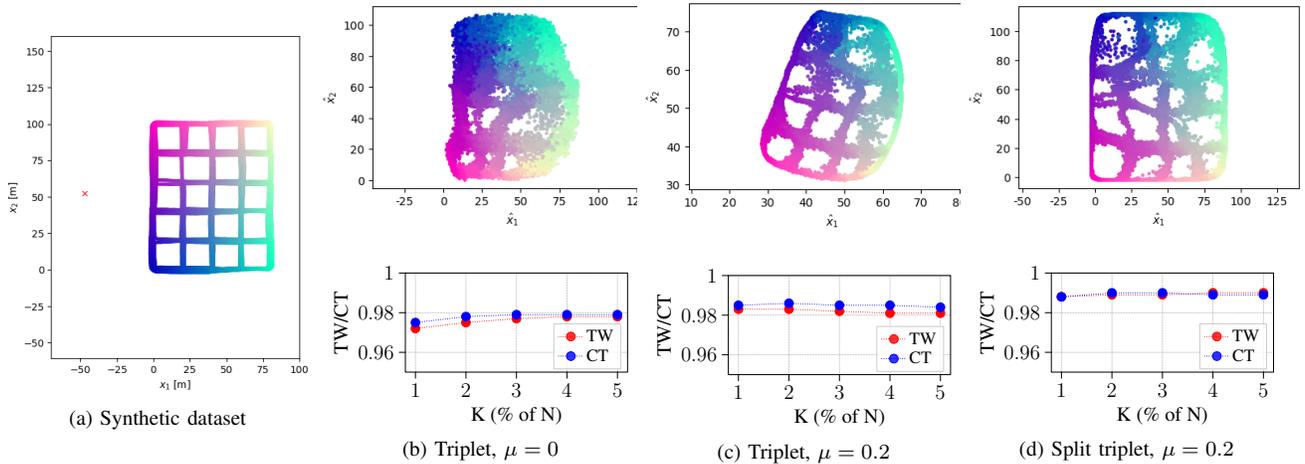
\begin{figure*}[tp]
\begin{minipage}{0.23\textwidth}
\centering
\subfloat[Synthetic dataset]{
\includegraphics[width=\textwidth]{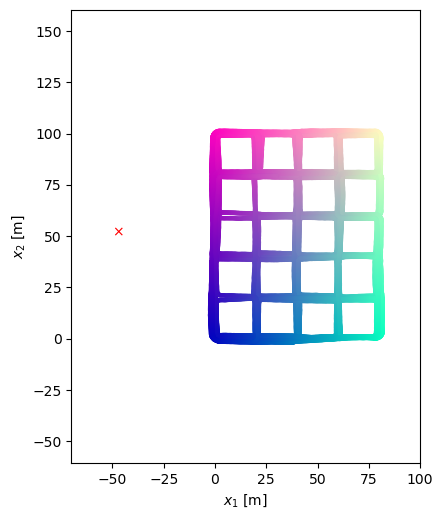}
\label{fig:scenario_figs_syn}
}
\end{minipage}
\begin{minipage}{0.23\textwidth}
\centering
\includegraphics[width=1.1\textwidth]{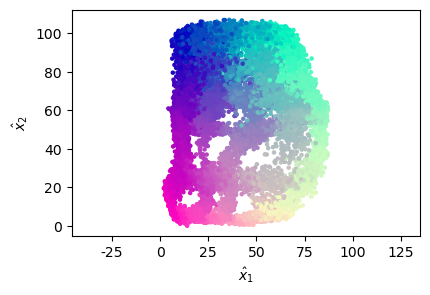}
\subfloat[Triplet, $\mu = 0$]{
\resizebox{\textwidth}{!}{
\begin{tikzpicture}

\begin{axis}[
legend cell align={left},
legend style={
  fill opacity=0.8,
  draw opacity=1,
  text opacity=1,
  at={(0.970,0.030)},
  anchor=south east,
  draw=white!80.000!black
},
tick align=outside,
tick pos=left,
x grid style={white!69.020!black},
xlabel={K (\% of N)},
xmajorgrids,
xmin=0.800, xmax=5.200,
xtick style={color=black},
y grid style={white!69.020!black},
ylabel={TW/CT},
ymajorgrids,
ymin=0.950, ymax=1.000,
ytick style={color=black}
]
\addplot [semithick, red, dotted, mark=*, mark size=3.000, mark options={solid}]
table {%
1.000 0.972
2.000 0.975
3.000 0.977
4.000 0.978
5.000 0.978
};
\addlegendentry{TW}
\addplot [semithick, blue, dotted, mark=*, mark size=3.000, mark options={solid}]
table {%
1.000 0.975
2.000 0.978
3.000 0.979
4.000 0.979
5.000 0.979
};
\addlegendentry{CT}
\draw (axis cs:0.050,0.050) node[
  scale=0.500,
  anchor=base west,
  text=black,
  rotate=0.0
]{KS = 0.182, SR = 0.080, SS = 0.269};
\end{axis}

\end{tikzpicture}
\label{fig:synth_trip_no_inertia}
}
}
\end{minipage}
\begin{minipage}{0.23\textwidth}
\centering
\includegraphics[width=1.1\textwidth]{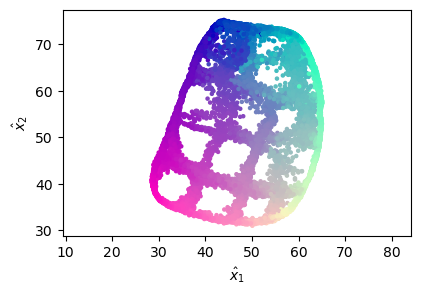}
\subfloat[Triplet, $\mu = 0.2$]{
\resizebox{\textwidth}{!}{
\begin{tikzpicture}

\begin{axis}[
legend cell align={left},
legend style={
  fill opacity=0.8,
  draw opacity=1,
  text opacity=1,
  at={(0.970,0.030)},
  anchor=south east,
  draw=white!80.000!black
},
tick align=outside,
tick pos=left,
x grid style={white!69.020!black},
xlabel={K (\% of N)},
xmajorgrids,
xmin=0.800, xmax=5.200,
xtick style={color=black},
y grid style={white!69.020!black},
ylabel={TW/CT},
ymajorgrids,
ymin=0.950, ymax=1.000,
ytick style={color=black}
]
\addplot [semithick, red, dotted, mark=*, mark size=3.000, mark options={solid}]
table {%
1.000 0.983
2.000 0.983
3.000 0.982
4.000 0.981
5.000 0.981
};
\addlegendentry{TW}
\addplot [semithick, blue, dotted, mark=*, mark size=3.000, mark options={solid}]
table {%
1.000 0.985
2.000 0.986
3.000 0.985
4.000 0.985
5.000 0.984
};
\addlegendentry{CT}
\draw (axis cs:0.050,0.050) node[
  scale=0.500,
  anchor=base west,
  text=black,
  rotate=0.0
]{KS = 0.158, SR = 0.048, SS = 0.210};
\end{axis}

\end{tikzpicture}
\label{fig:synth_trip_inertia}
}
}
\end{minipage}
\begin{minipage}{0.23\textwidth}
\centering
\includegraphics[width=1.1\textwidth]{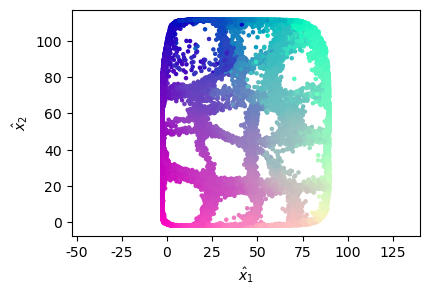}
\subfloat[Split triplet, $\mu = 0.2$]{
\resizebox{\textwidth}{!}{
\begin{tikzpicture}

\begin{axis}[
legend cell align={left},
legend style={
  fill opacity=0.8,
  draw opacity=1,
  text opacity=1,
  at={(0.970,0.030)},
  anchor=south east,
  draw=white!80.000!black
},
tick align=outside,
tick pos=left,
x grid style={white!69.020!black},
xlabel={K (\% of N)},
xmajorgrids,
xmin=0.800, xmax=5.200,
xtick style={color=black},
y grid style={white!69.020!black},
ylabel={TW/CT},
ymajorgrids,
ymin=0.950, ymax=1.000,
ytick style={color=black}
]
\addplot [semithick, red, dotted, mark=*, mark size=3.000, mark options={solid}]
table {%
1.000 0.988
2.000 0.989
3.000 0.989
4.000 0.990
5.000 0.990
};
\addlegendentry{TW}
\addplot [semithick, blue, dotted, mark=*, mark size=3.000, mark options={solid}]
table {%
1.000 0.988
2.000 0.990
3.000 0.990
4.000 0.989
5.000 0.989
};
\addlegendentry{CT}
\draw (axis cs:0.050,0.050) node[
  scale=0.500,
  anchor=base west,
  text=black,
  rotate=0.0
]{KS = 0.119, SR = 0.028, SS = 0.160};
\end{axis}

\end{tikzpicture}
\label{fig:synth_strip_inertia}
}
}
\end{minipage}
\caption{Top: learned channel charts from the synthetic dataset (locations shown in \fref{fig:scenario_figs_syn}); bottom: TW/CT dependent on neighborhood size $K$ (\% of $N = 17\,769$ CSI vectors). Left: conventional triplet loss as in~\cite{huawei2020channelcharting}; middle: triplet loss with inertial regularizer ($\mu = 0.2$); right: proposed split triplet loss with inertial regularizer ($\mu = 0.2$). We observe that (i) inertial regularization significantly improves the channel chart and (ii) the proposed split triplet loss performs on par with the conventional triplet loss in terms of TW/CT but more resemble real locations as they occupy the entire area.}
\label{fig:synth}
\end{figure*}

\subsection{Triplet Selection}
\label{sec:choice_of_triplets}

As mentioned in \fref{sec:triplet}, triplets can be selected based solely on time stamps. In~\cite{huawei2020channelcharting}, an anchor point $i$ is chosen at random from the dataset, at time stamp $t_i$; then the positive point $j$ is chosen from $\{j: 0 < |t_i - t_j| < T_c\}$ and the negative point $k$ is chosen from $\{k: T_c < |t_i - t_k| < T_f\}$, where $T_c$ and $T_f$ are user-defined parameters. 
Although intuitive, this approach cannot correctly identify self-intersecting tracks. If a UE crosses its own path from some time prior, points could be arbitrarily far in time but close in distance.

To address this problem, we return, somewhat surprisingly, to the feature distance.
As mentioned before, feature distance is not generally a suitable proxy for real distance, but if the feature distance is close to zero, then the true distance is typically small.
We leverage this observation as a way to detect self-intersections: if a point is far in time but close in feature distance (as measured using a quantile threshold over all feature distances), then it is likely to be a self-intersection point and should be treated as a nearby point instead.

\subsection{Inertial Regularization}
\label{sec:inertia}

Throughout this paper, we have been guided by the principle that we can and should use all information we have available to us. In a physical system, we can make a simple observation: the track of a body moving through real space is smooth. 
By this we mean that the UE's path must be continuous, so it cannot jump far away in a short time; this corresponds to the temporally-based triplet selection mechanism mentioned in \fref{sec:choice_of_triplets}. But we can gleam additional information from this assertion: a body also cannot change its velocity discontinuously, which would imply infinite acceleration. 
In order to make use of this information, we propose a new regularizer which captures this inertia of the UE. The regularizer, which we unsurprisingly call the \emph{inertial regularizer}, takes the following form:
\begin{align}
\mathcal{L}_{\text{Inertia}}(\bithetad) \define
\sum_{(i,j,\ell)\in\mathcal{I}}\|(\hat{\bmx}_j^\bithetad - \hat{\bmx}_i^\bithetad) - (\hat{\bmx}_i^\bithetad - \hat{\bmx}_\ell^\bithetad)\|,
\end{align}
where $\mathcal{I} = \{(i,j,\ell) : t_i - t_\ell = t_j - t_i, |t_i - t_j| < T_c\}$, temporally linear triplets. The regularizer is the second difference of consecutive points: if $i$, $j$, and $\ell$ are equally spaced and on a straight line, then it will be zero, but it penalizes the UE slowing down, speeding up, or turning.
In real motion, with a high enough sampling rate, the UEs move linearly or with gradual turns or changes in acceleration.
Forcing this regularizer to be zero would result in linear, evenly spaced points. Instead, we deweight it using a suitably chosen regularization factor~$\mu$. We can modify the existing temporal triplet selection procedure to provide inertial points: for an anchor point $i$ with positive point $j$, the inertial point is at time stamp $t_i - (t_j - t_i)$, for points close in time; we ignore inertial loss on points re-identified by feature distance (see \fref{sec:choice_of_triplets} for the details).


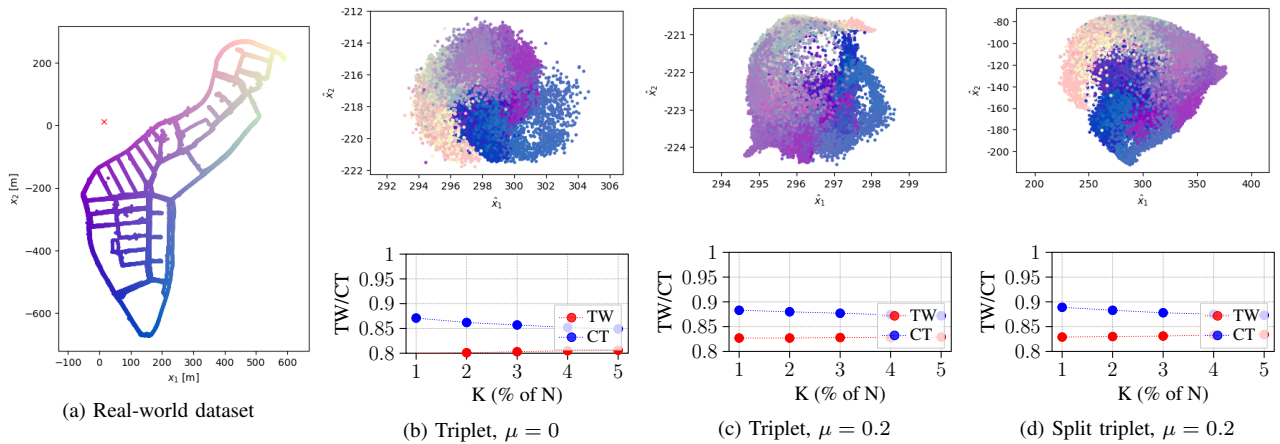
\begin{figure*}[tp]
\centering
\begin{minipage}{0.23\textwidth}
\centering
\subfloat[Real-world dataset]{
\includegraphics[width=\textwidth]{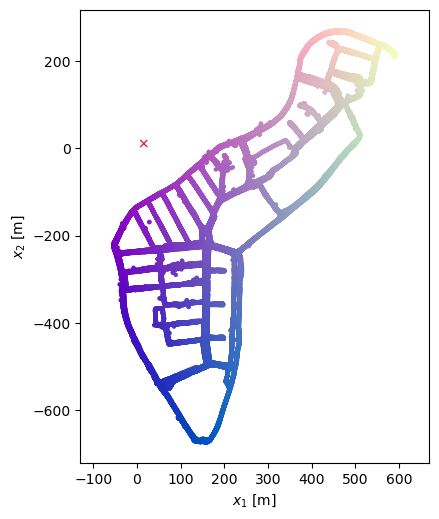}
\label{fig:scenario_figs_real}
}
\end{minipage}
\begin{minipage}{0.23\textwidth}
\centering
\includegraphics[width=\textwidth]{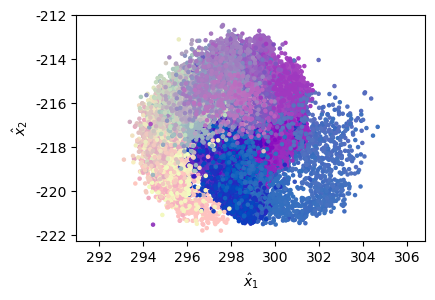}
\subfloat[Triplet, $\mu = 0$]{
\resizebox{\textwidth}{!}{
\begin{tikzpicture}

\begin{axis}[
legend cell align={left},
legend style={
  fill opacity=0.8,
  draw opacity=1,
  text opacity=1,
  at={(0.970,0.030)},
  anchor=south east,
  draw=white!80.000!black
},
tick align=outside,
tick pos=left,
x grid style={white!69.020!black},
xlabel={K (\% of N)},
xmajorgrids,
xmin=0.800, xmax=5.200,
xtick style={color=black},
y grid style={white!69.020!black},
ylabel={TW/CT},
ymajorgrids,
ymin=0.800, ymax=1.000,
ytick style={color=black}
]
\addplot [semithick, red, dotted, mark=*, mark size=3.000, mark options={solid}]
table {%
1.000 0.799
2.000 0.801
3.000 0.803
4.000 0.805
5.000 0.806
};
\addlegendentry{TW}
\addplot [semithick, blue, dotted, mark=*, mark size=3.000, mark options={solid}]
table {%
1.000 0.871
2.000 0.862
3.000 0.857
4.000 0.852
5.000 0.849
};
\addlegendentry{CT}
\draw (axis cs:0.050,0.050) node[
  scale=0.500,
  anchor=base west,
  text=black,
  rotate=0.0
]{KS = 0.464, SR = 0.841};
\end{axis}

\end{tikzpicture}
\label{fig:real_trip_no_inertia}
}
}
\end{minipage}
\begin{minipage}{0.23\textwidth}
\centering
\includegraphics[width=\textwidth]{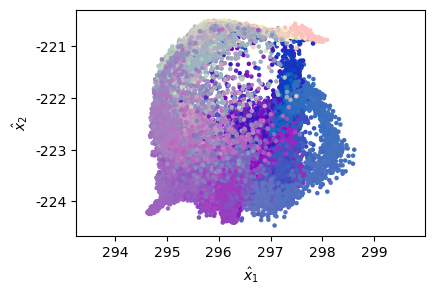}
\subfloat[Triplet, $\mu = 0.2$]{
\resizebox{\textwidth}{!}{
\begin{tikzpicture}

\begin{axis}[
legend cell align={left},
legend style={
  fill opacity=0.8,
  draw opacity=1,
  text opacity=1,
  at={(0.970,0.030)},
  anchor=south east,
  draw=white!80.000!black
},
tick align=outside,
tick pos=left,
x grid style={white!69.020!black},
xlabel={K (\% of N)},
xmajorgrids,
xmin=0.800, xmax=5.200,
xtick style={color=black},
y grid style={white!69.020!black},
ylabel={TW/CT},
ymajorgrids,
ymin=0.800, ymax=1.000,
ytick style={color=black}
]
\addplot [semithick, red, dotted, mark=*, mark size=3.000, mark options={solid}]
table {%
1.000 0.827
2.000 0.827
3.000 0.828
4.000 0.828
5.000 0.829
};
\addlegendentry{TW}
\addplot [semithick, blue, dotted, mark=*, mark size=3.000, mark options={solid}]
table {%
1.000 0.883
2.000 0.880
3.000 0.877
4.000 0.874
5.000 0.872
};
\addlegendentry{CT}
\draw (axis cs:0.050,0.050) node[
  scale=0.500,
  anchor=base west,
  text=black,
  rotate=0.0
]{KS = 0.473, SR = 0.828};
\end{axis}

\end{tikzpicture}
\label{fig:real_trip_inertia}
}
}
\end{minipage}
\begin{minipage}{0.23\textwidth}
\centering
\includegraphics[width=\textwidth]{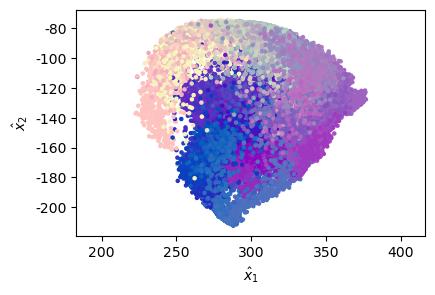}
\subfloat[Split triplet, $\mu = 0.2$]{
\resizebox{\textwidth}{!}{
\begin{tikzpicture}

\begin{axis}[
legend cell align={left},
legend style={
  fill opacity=0.8,
  draw opacity=1,
  text opacity=1,
  at={(0.970,0.030)},
  anchor=south east,
  draw=white!80.000!black
},
tick align=outside,
tick pos=left,
x grid style={white!69.020!black},
xlabel={K (\% of N)},
xmajorgrids,
xmin=0.800, xmax=5.200,
xtick style={color=black},
y grid style={white!69.020!black},
ylabel={TW/CT},
ymajorgrids,
ymin=0.800, ymax=1.000,
ytick style={color=black}
]
\addplot [semithick, red, dotted, mark=*, mark size=3.000, mark options={solid}]
table {%
1.000 0.829
2.000 0.830
3.000 0.831
4.000 0.833
5.000 0.834
};
\addlegendentry{TW}
\addplot [semithick, blue, dotted, mark=*, mark size=3.000, mark options={solid}]
table {%
1.000 0.889
2.000 0.883
3.000 0.878
4.000 0.875
5.000 0.873
};
\addlegendentry{CT}
\draw (axis cs:0.050,0.050) node[
  scale=0.500,
  anchor=base west,
  text=black,
  rotate=0.0
]{KS = 0.423, SR = 0.824};
\end{axis}

\end{tikzpicture}
\label{fig:real_strip_inertia}
}
}
\end{minipage}
\caption{Top: learned channel charts from the real-world IEEE CTW 2020 data competition dataset (locations shown in \fref{fig:scenario_figs_real}); bottom: TW/CT dependent on neighborhood size $K$ as a percentage of $N = 45\,784$ CSI vectors. Left: basic triplet loss as in~\cite{huawei2020channelcharting}; middle: triplet loss with inertial regularizer ($\mu=0.2$); right: proposed split triplet loss with inertial regularizer ($\mu=0.2$). We note that (i) inertial regularization significantly improves the channel chart and (ii) the proposed split triplet loss outperforms the basic triplet loss in terms of TW/CT and more closely resemble real locations as they occupy the full area.}
\label{fig:real}
\end{figure*}

\section{Results}
\label{sec:results}

We now provide results on synthetic and real-world measured CSI datasets: first, we introduce performance metrics and summarize the neural network structure used in the algorithms; second, we present our results compared to two benchmarks, Sammon's mapping via Siamese nets \cite{ericpaper} and the triplet loss~\cite{huawei2020channelcharting}.

\subsection{Performance Metrics}

Channel charting is inherently a self-supervised learning problem, which means that standard error metrics, such as the mean-squared error, are not applicable. Instead, we consider the following performance metrics to assess channel chart quality.

\subsubsection{Kruskal Stress}
Kruskal stress~\cite{kruskalstress} is commonly used in multidimensional scaling, the problem of finding a low-dimensional set of points which preserve distances in the high-dimensional space. We use the following definition \cite{ericpaper}:
\begin{align}
\label{eq:Kruskal}
\textit{KS} \define \textstyle \sqrt{1 - \left(\frac{\langle\bmd, \hat{\bmd}\rangle}{\|\bmd\|\cdot\|\hat{\bmd}\|}\right)^2}.
\end{align}
Here, $\bmd$ and $\hat{\bmd}$ are vectorized versions of the pairwise distance matrices formed by $\{d_{ij}\}$ and $\{\hat{d}_{ij}\}$; lower is better. This metric can also be interpreted in terms of the cosine distance between~$\bmd$ and~$\hat{\bmd}$: if $\frac{\langle \bmd, \hat{\bmd}\rangle}{\|\bmd\|\cdot\|\hat{\bmd}\|} = \cos\phi$, then $\textit{KS} = \sin\phi$.
The Kruskal stress in \fref{eq:Kruskal} is independent of scale, rotation, or translation, since it is measured using normalized distances. It is a useful similarity measurement for global distances, but does not specifically measure local embedding quality.

\subsubsection{Optimal Shifting, Stretching, and Rotating}

Since the channel charts are created using relative distances, they are invariant to translations and rotations. In addition,  there cannot be a measure of absolute scale since the distances are given by a partial ordering. Accordingly, the optimal shifting, stretching, and rotating metric (SR) \cite{huawei2020channelcharting} is invariant to these parameters:
\begin{align}
\textit{SR} \define \frac{1}{N}\sum_i \left\|\bD_{\bm{\sigma}_{x}}^{-1}(\bmx_i-\bmmu_x) - \bW\bD_{\bm{\sigma}_{\hat{x}}}^{-1}(\hat{\bmx}_i - \bmmu_{\hat{x}})\right\|^2.
\end{align}
Here, $\bmmu_x = \frac{1}{N}\sum_i\bmx_i$, $\bD_{\bm{\sigma}_x} = \text{diag}((\frac{1}{N}\sum_i(\bmx_i - \bmmu_x)^{\circ 2})^{\circ 1/2})$, $\bmmu_{\hat{x}}$ and $\bD_{\bm{\sigma}_{\hat{x}}}$ are defined analogously, and $\bW$ is the optimal projection from normalized embeddings onto normalized ground truth solving the orthogonal Procrustes problem, $\bW = \bU\bV^T$ with  the singular value decomposition $\bU\bS\bV^T$ of $(\bD_{\bm{\sigma}_{\hat{x}}}^{-1}(\hat{\bX} - \vecone\bmmu_{\hat{x}}^T))^T\bD_{\bm{\sigma}_{x}}^{-1}(\bX-\vecone\bmmu_x^T)$, where $\bX$ and $\hat{\bX}\in\reals^{N\times D}$ are the matrices created by concatenating the $N$ ground truth and learned locations, respectively \cite{Schonemann1966AGS}.

\subsubsection{Trustworthiness and Continuity}

Local embedding quality uses a different set of metrics. Trustworthiness (TW) captures whether the neighbors in the embedding are also neighbors in the ground truth; continuity (CT) measures the converse~\cite{ericpaper}.
Each is parametrized over a neighborhood of size $K$; we consider values of $K$ from 1\% to 5\% of the total number of samples, $N$. If the set of $K$ closest neighbors by Euclidean distance to a point $i$ in the embedding is $\mathcal{U}_K(i)$, then
\begin{align}
\textit{TW}(K) \define 1 - \tfrac{1}{(2N - 3K - 1)NK}\sum_i\sum_{j\in\mathcal{U}_K(i)}(r(i,j) - K),
\end{align}
where $r(i,j)$ is the order of $j \ne i$, ranked by distance from $i$ in the ground truth. Similarly, if $\mathcal{V}_K(i)$ is the set of $K$ closest neighbors to $i$ in the ground truth, then
\begin{align}
\textit{CT}(K) \define 1 - \tfrac{1}{(2N - 3K - 1)NK}\sum_i\sum_{j\in\mathcal{V}_K(i)}(\hat{r}(i,j) - K),
\end{align}
where $\hat{r}(i,j)$ is the analogous rank ordering in the embeddings; the leading coefficient is a normalization factor which ensures that $\textit{TW}(K)$ and $\textit{CT}(K)$ are between $0$ and $1$; higher is better.

\subsection{Neural Network Structure}
In the following experiments, we use the same  neural network architecture for the function $g_\bithetad$ that maps CSI features to points in the channel chart. 
We use a neural network with $2048$ inputs, three dense layers, halving the number of nodes in each subsequent layer, and $256$ outputs. We use ReLU activations except for the last layer, which generates probability maps~\cite{emrepaper} and uses a softmax function instead. This final layer outputs a probability mass function (PMF) $\bmp_i = \tilde{g}_\bithetad(\bmf_i)$ over a set of predefined grid centroids $\{\bmc_k\}$ with $\bmc_k\in\reals^{D'}$ which span the area, instead of predicting the pseudo-location in $\reals^{D'}$ directly; $p_k$ indicates the likelihood of the UE being at grid point $c_k$. We can then calculate the expected pseudo-location using the PMF and the $16\times 16$ set of centroids, $\hat{\bmx}_i = \sum_{k}\bmc_k p_k$.

\subsection{Synthetic Results}

We evaluate the performance of our methods on a synthetic dataset with CSI generated using the QuadRiGa urban macro non-line-of-sight Berlin scenario \cite{quadriga}. 
We created a simple city-block MATLAB simulator in which a mobile UE travels randomly around a $4\times 4$ grid of blocks.
Using this simulator, we generated a dataset of $N = 17\,769$ channel vectors using a $B = 32$ antenna BS with a ULA and half wavelength antenna spacing. Thermal noise was added to the CSI vectors dependent on room temperature (300~K) and the bandwidth, and the signals were received using an OFDM system with $W = 64$ subcarriers at a bandwidth of 20\,MHz. The BS is located around 50\,m away from the UE locations; see \fref{fig:scenario_figs_syn} for further details.

As shown in \fref{fig:synth}, both algorithms are able to convincingly approximate the dataset. In \fref{fig:synth_trip_no_inertia}, we compare the original algorithm from \cite{huawei2020channelcharting}, the triplet loss with a margin $\lambda = 1$ and no inertia: this algorithm is able to recreate some of the city-block topology but contains irregularities. In \fref{fig:synth_trip_inertia}, we add the inertial regularizer described in \fref{sec:inertia} to the triplet loss, which improves the local accuracy but negatively affects the scaling. Finally, in \fref{fig:synth_strip_inertia}, we demonstrate the split triplet loss with the inertial regularizer, which has trustworthiness and continuity similar to that of the triplet loss but lower Kruskal stress and optimal stretch/rotation; see Table \ref{table:syn} for the details.

\begin{table}[tp]
\centering
\caption{Results for synthetically generated non-LoS data.}
\label{table:syn}
\begin{tabular}{@{}lc|cccc@{}}
\toprule
Method & $\mu$ & $\textit{KS}$ & $\textit{SR}$ & $\textit{TW}$ (5\%) & $\textit{CT}$ (5\%) \\ 
\midrule
\multirow{2}{*}{Sammon's mapping} & 0 & 0.525 & 1.132 & 0.647 & 0.731 \\
& 0.2 & 0.511 & 1.063 & 0.665 & 0.749 \\
\midrule
\multirow{2}{*}{Triplet loss} & 0 & 0.182 & 0.080 & 0.978 & 0.979 \\
& 0.2 & 0.158 & 0.048 & 0.981 & 0.984 \\
\midrule
\multirow{2}{*}{Split triplet loss} & 0 & 0.200 & 0.085 & 0.956 & 0.956 \\
& 0.2 & {\bf 0.119} & {\bf 0.028} & {\bf 0.990} & {\bf 0.989} \\
\bottomrule
\end{tabular}
\end{table}

\subsection{Measured Results}
We also demonstrate the performance of our methods on a real-world dataset consisting of channel matrices from the published dataset from \cite{arnold2019novel} of the IEEE Communication Theory Workshop (CTW) 2020 data competition. The dataset covers a large suburban area (see \fref{fig:scenario_figs_real}). The UE in the field is equipped with a differential global positioning system (GPS), creating tagged CSI with 20\,MHz of bandwidth at a carrier frequency of 1.27\,GHz. The BS is equipped with a URA consisting of $8\times8$ antennas mounted on a high building complex. The measurement sanity was verified for massive MIMO precoding, thus capturing the spatial component of the environment correctly.
Data was collected using SIMO-OFDM with $W = 924$ subcarriers for about 8 hours, resulting in $N = 45\,784$ CSI measurements. We calculated CSI features and averaged over 5 independent measurements.

It is harder to learn accurate embeddings on the real dataset, since it is larger and more sparsely sampled in space than the synthetic dataset, and the performance is correspondingly lower, as seen in \fref{fig:real}. We observe the same patterns of behavior as in the synthetic dataset. The benchmark algorithm with the triplet loss and no inertial regularization in \fref{fig:real_trip_no_inertia} is able to separate points roughly by location, but is poorly scaled and misshapen. When we add the inertia regularizer, \fref{fig:real_trip_inertia} shows slightly improved performance (it is able to separate the blue ``elbow'' of points at the bottom of the ground truth around (150,-600) in \fref{fig:scenario_figs_real} from the other points, for example) but still suffers from overlapping points and a bad scale. Finally, in \fref{fig:real_strip_inertia} we show the triplet loss combined with the inertial regularizer, which is able to outperform the other methods, once again marginally in terms of SR, TW, and CT and more so in Kruskal stress; see Table \ref{table:real} for the details.

\begin{table}[tp]
\centering
\caption{Results for the measured IEEE CTW 2020 dataset.}
\label{table:real}
\begin{tabular}{@{}lc|cccc@{}}
\toprule
Method & $\mu$ & $\textit{KS}$ & $\textit{SR}$ & $\textit{TW}$ (5\%) & $\textit{CT}$ (5\%) \\ 
\midrule
\multirow{2}{*}{Sammon's mapping} & 0 & 0.610 & 1.564 & 0.540 & 0.576 \\
& 0.2 & 0.582 & 1.433 & 0.619 & 0.669 \\
\midrule
\multirow{2}{*}{Triplet loss} & 0 & 0.464 & 0.841 & 0.806 & 0.849 \\
& 0.2 & 0.473 & 0.828 & 0.829 & 0.872 \\
\midrule
\multirow{2}{*}{Split triplet loss} & 0 & 0.459 & 0.877 & 0.786 & 0.831 \\
& 0.2 & {\bf 0.423} & {\bf 0.824} & {\bf 0.834} & {\bf 0.873} \\
\bottomrule
\end{tabular}
\end{table}


\section{Conclusions}
\label{sec:conclusions}

We have proposed the split triplet loss for channel charting, which takes into account physical constraints from the CSI measurement process, as well as a suitable triplet selection procedure and a novel inertial regularizer that incorporates the fact that moving UEs must have finite acceleration/deceleration. 
We have demonstrated the efficacy of our methods on a real-world dataset, which reveals that our methods outperform the Siamese network of \cite{ericpaper} and the state-of-the-art triplet loss of \cite{huawei2020channelcharting} in terms of Kruskal stress, optimal stretching/rotating, trustworthiness, and continuity. 
We leave several problems as future work: an obvious extension is to utilize multiple BSs/APs as well as semi-supervised techniques as in \cite{ericpaper} to enable absolute positioning capabilities with channel charting. 

\section*{Acknowledgments}
The authors would like to thank Olav Tirkkonen and Pengzhi Huang for discussions on channel charting. 

\balance

\bibliographystyle{IEEEtran}
\bibliography{VIPabbrv,jsacbib}

\balance

\end{document}